\documentclass[prd,preprint,nofootinbib]{revtex4-1} 

\usepackage{latexsym,verbatim}
\usepackage{amssymb}
\usepackage{amsfonts}
\usepackage{amsmath,color}
\usepackage[draft]{graphicx}
\usepackage{bm}

\usepackage{hyperref}

\hypersetup{
    colorlinks=true,
    linkcolor=red,
    citecolor=green,
    filecolor=magenta,      
    urlcolor=cyan,
    pdftitle={Galactic Dynamics in General Relativity: the Role of Gravitomagnetism},
    pdfpagemode=FullScreen,}

\def\mb#1{\mathbf{#1}}

\def\ber{\begin{eqnarray}}
\def\eer{\end{eqnarray}}
\def\beq{\begin{equation}}
\def\eeq{\end{equation}}

\def\rmd{{\rm d}}

\def\ed{\end{document}}

\newcommand{\ppar}[2]{\frac{\partial #1}{\partial #2}}

\begin{document}

\author{Davide Astesiano}
\email{davide.astesiano@yahoo.it}
\affiliation{Universit\`a degli studi dell'Insubria, Dipartimento di Scienza ed Alta Tecnologia, Via Valleggio 11, 22100 Como - Italy}
\affiliation{INFN - Via Celoria 16, 20133 Milano, Italy}

\author{Matteo Luca Ruggiero}
\email{matteoluca.ruggiero@unito.it}
\affiliation{Dipartimento di Matematica ``G.Peano'', Universit\`a degli studi di Torino, Via Carlo Alberto 10, 10123 Torino, Italy}
\affiliation{INFN - LNL , Viale dell'Universit\`a 2, 35020 Legnaro (PD), Italy}

\date{\today}

\title{Galactic dark matter effects from purely geometrical aspects of General Relativity}
\begin{abstract}
We study  disc galaxies in the framework of General Relativity to focus on the possibility that,  even in the low energy limit, there are relevant corrections with respect to the purely Newtonian approach. Our analysis encompasses the model both considering a low energy expansion  and exact solutions, making it clear the connection between these different approaches. In particular, we focus on two different limits: the well known gravitomagnetic analogy and a new limit, called ``strong gravitomagnetism'', which has corrections in $c$ of the same order as the Newtonian terms. We show that these two limits of the general class of solutions can account for the observed flat velocity profile, contrary to what happens using Newtonian models, where a dark matter contribution is required. Hence, we suggest a geometrical origin for a certain amount of dark matter effects.
\end{abstract}

\maketitle

\section{Introduction}\label{sec:intro}

One of the evidences supporting the existence of \textit{dark matter} comes from the observations of the rotation curves of galaxies, which are flat contrary to what is expected on the basis of Newtonian dynamics \cite{strigari2013galactic}. In this context, Newtonian gravity rather than General Relativity (GR)  is used because far from the galactic center (where the flat behaviour is observed) the gravitational field is reasonably supposed to be weak and stars are not moving at relativistic speeds. 
Nonetheless, it was conjectured that GR may have a role in this context: in particular, the problem of galactic rotation curves was studied both considering exact solutions of GR equations \cite{Cooperstock:2006dt,Balasin:2006cg,crosta2020testing,Astesiano:2021ren} and weak-field approximations \cite{Ramos-Caro:2012ren,ludwig2021galactic,Ruggiero:2021lpf}.  Indeed,  there are general relativistic effects without a Newtonian analogue, such as the gravitomagnetic effects, deriving from mass currents. In the above-cited papers, using different approaches, it was  suggested that if these non-Newtonian effects were taken into account, the impact of dark matter in explaining the observations could be different. 

The purpose of this paper is to focus on the role that gravitomagnetic and, more in general, post-Newtonian effects might have in galactic dynamics. In order to trace the impact of these effects from a very general viewpoint, we will not resort to a specific galaxy model, but we will emphasize the modifications introduced by general relativistic effects starting from very few hypotheses, that basically refer to the underlying symmetries.\\ 
In particular,  in Section \ref{Gravitomagn} we describe how the weak-field approach to the solution of Einstein's equations, which leads to the well know gravitoelectromagnetic analogy \cite{Ruggiero:2002hz,Mashhoon:2003ax}, can be used to investigate the possible impact of GR effects on  galactic rotation curves.
Subsequently, in Section \ref{GRS} we focus on the exact general relativistic solutions for an axisymmetric stationary system coupled to dust \cite{Astesiano:2021ren}, discussing its physical properties and the relevant limits. A new weak-field limit of the general solution that we call  ``strong gravitomagnetism'' (SGM)  is introduced in Section \ref{GEM}:  we suggest that it can provide an interesting model for disc galaxies. In this regards, we compare this limit with the Newtonian one (Section \ref{Newtonian}) and we highlight the differences: unlike the Newtonian model, the SGM limit can naturally provide a flat velocity curve and the presence of a non-diagonal term in the metric can  reduce the amount of energy-density needed to sustain the motion of the galaxy.\\
Eventually, in Section \ref{Rigid} we consider  the rigidly rotating solution: it coincides with the Balasin-Grumiller model \cite{Balasin:2006cg}, which recently gained relevance since, starting from this model,  \citet{crosta2020testing} showed a good agreement between the model and the Gaia\cite{GAIA1,GAIA2} data for the Milky way. We show that this model, being a rigidly rotating solution, presents some unphysical features which need to be addressed: for instance, the redshift (or blueshift) due to the emission of photons from the galaxy measured from an asymptotic inertial observer is linearly increasing.\\
{\indent Before starting with the technical discussion, we point out that the role of gravitomagnetism, and also of the other orders in $c^{-n}$, is usually understood in vacuum, i.e. outside of the source. In this configuration it makes sense to say that there is a dominant Newtonian contribution plus corrective terms of higher order in $c^{-n}$. What we do here is substantially different and the usually adopted expansion in vacuum breaks down. This is due to the fact that we are not analyzing the equations in vacuum, but "inside" the matter distribution. i.e. within the galaxy. In Section \ref{Gravitomagn} we will stress again this important difference.}

\section{Gravitomagnetic effects in galaxies}  \label{Gravitomagn}

It possible to write the solution of Einstein's field equations in weak-field and slow-motion approximation exploiting a well known analogy with Maxwell equations: this is the so-called \textit{gravitoelectromagnetic} formalism (see e.g. \citet{Ruggiero:2002hz,Mashhoon:2003ax}); accordingly, the line element describing this solution is
\beq
\mathrm{d} s^2= -c^2 \left(1-2\frac{\Phi}{c^2}\right)\rmd t^2 -\frac4c ({\mathbf A}\cdot \rmd {\mathbf x})\rmd t + \left(1+2\frac{\Phi}{c^2}\right)\delta_{ij}\rmd x^i \rmd x^j. \label{eq:weakfieldmetric1}
\eeq
In the above equation the gravitoelectric  ($\Phi $) and gravitomagnetic  ($\mathbf A$) potentials, in stationary conditions, are solutions of the Poisson equations 
\begin{eqnarray}
\nabla^{2} \Phi&=&-4\pi G\, \rho, \label{eq:poisson01} \\
\nabla^{2} \mb A & =& -\frac{8\pi G}{c}\, \mb j,  \label{eq:poisson02}
\end{eqnarray}
in terms of the mass density $\rho_{  }$ and current $\mathbf j$ of the sources. 

Notice that, in the gravitoelectromagnetic formalism, in analogy with the electric potential of a point charge,  $\Phi$ differs by a minus sign from the actual Newtonian potential of point mass $M$, $U=-\frac{GM}{|\mb x|}$, which we use in Section \ref{sec:limits}.

Starting from the above potentials, in stationary conditions, we may define the gravitoelectric ($\mb E$) and gravitomagnetic ($\mb B$) fields 
\begin{eqnarray}
\mb E&=&-\bm \nabla \Phi, \label{eq:defE} \\
\mb B&=& \bm \nabla \times \mb A. \label{eq:defB}
\end{eqnarray}
Using these fields, Einstein's equations can be written in analogy to Maxwell's equations. In addition, the spatial component of the geodesic equation (up to linear order in $\frac{|\mb v|}{c}$) is written in terms of  Lorentz-like force  acting upon a test mass $m$
\beq
m\frac{\rmd {\mathbf v}}{\rmd t}=-m{\mathbf E}-2m \frac{{\mathbf v}}{c}\times {\mathbf B}. \label{eq:lor2}
\eeq
This formalism is useful since it allows to express GR effects in terms of known electromagnetic ones: for instance, the 
Lense-Thirring effect can be explained in analogy with the precession of a magnetic dipole in a magnetic field (see e.g. \citet{iorio2011}). However, this formalism has limitations (for instance  the geodesic equation does not take a Lorentz-like form in non stationary conditions as discussed by \citet{Ruggiero:2021uag}) and we should not forget that it is just an approximation of the full theory.\\ 
In the context of the study of galactic dynamics,  \citet{ludwig2021galactic} considered the set of gravitational equations for a fluid of stars, modelled as dust: in particular he solved, in stationary conditions, the momentum equation (\ref{eq:lor2}) and the source equations 
( \ref{eq:poisson01})-(\ref{eq:poisson02}) to obtain a self-consistent solutions for  $\mb v, \mb A, \Phi$, and showed that the impact of gravitomagnetic effects on the rotation curves is not negligible.

Without using a specific model for the density profile of a galaxy, it is possible to deduce that gravitomagnetic effects may have a relevant impact on the galactic rotation curves, as discussed by \citet{Ruggiero:2021lpf}. To this end, we consider dust particles  steadily rotating around a symmetry axis and use cylindrical coordinates $\{r,\varphi,z\}$  such that $z$ is the rotation axis; $\mb u_{r}, \mb u_{\varphi}, \mb u_{z}$ are the unit vectors. If $\bm \Omega=\Omega \mb u_{z}$ is the rotation rate and $\mb x$ is the position vector of a dust particle, its velocity turns out to be $\displaystyle \mb v=\bm \Omega \times \mb x$, and $\bm \Omega$ can be a function of ${r}$ and ${z}$, since axial symmetry is assumed. Accordingly, using a purely Newtonian model in stationary condition the Poisson equation can be written as
\beq
4\pi G \rho=-\bm \nabla \cdot \left[\left(\mb v \cdot \bm \nabla \right)\mb v\right], \label{eq:rhoomega1}
\eeq
where $\mb v$ is the velocity field of the fluid: taking into account that $\mb v=\Omega r \mb u_{\varphi}=v \mb u_{\varphi} $, from Eq.  (\ref{eq:rhoomega1}) we obtain
\beq
4\pi G \rho=2\Omega^{2}+2\Omega\ppar{\Omega}{r}r=\frac{2 v}{r}\ppar{v}{r}. \label{eq:rhoomega}
\eeq
In this equation, the matter density $\rho$ is locally related to the rotation rate $\Omega$ and its derivative. If we focus on the regime where the rotation curves are flattened, since it is  $v=\Omega r \simeq \mathrm{constant}$, from (\ref{eq:rhoomega}) we get $\rho=0$: accordingly, using a Newtonian approach, it is not clear how to link the matter density to the rotation rate in the flat zone.

Things are quite different if we work in a GR context, in weak-field and slow-motion approximation. Indeed, exploiting the above described analogy with electromagnetism, Eq. (\ref{eq:rhoomega}) becomes
\beq
4\pi G \rho+\frac{2}{c}\mb B \cdot \bm \omega=-\bm \nabla \cdot \left[\left(\mb v \cdot \bm \nabla \right)\mb v\right]. \label{eq:poissmod}
\eeq
In the above equation $\bm \omega=\bm \nabla \times \mb v$ is the Newtonian fluid vorticity. Accordingly, the coupling between the gravitomagnetic field and the fluid vorticity modifies the local relation between density and the velocity of the fluid. If we set $\rho=\rho_{N}$ in Eq. (\ref{eq:rhoomega}) to denote the density measured in a Newtonian framework and  $\rho=\rho_{N} + \delta \rho$ in Eq. (\ref{eq:poissmod}), where $\delta \rho$ is the extra density due to the coupling between the gravitomagnetic field and the fluid vorticity, we obtain
\beq
 \delta \rho=-\frac{1}{2\pi G c}\mb B \cdot \bm \omega. \label{eq:posissondelta}
\eeq
This approach shows that, even in weak-field conditions, neglecting post-Newtonian effects might have an impact on the estimate of the mass density and, in turn, this could contribute to a different evaluation of the dark matter content. 

The impact on the evaluation of the matter content can be deduced also by the extension of the virial theorem in the gravitoelectromagnetic case, studied by \citet{Astesiano:2022gph}
\begin{align}
 \langle \int \rho v^2\, d^3x& - \frac{1}{2} \int \rho  \Phi\, d^3x - \frac{1}{8 \pi G} \int \mathcal{H} d^3x \rangle=0,\\   
    \mathcal{H}:=& \left(\partial_{\hat{i}} A_{\hat{j}}\right)^2- \left(\partial_{\hat{i}} A_{\hat{j}}\right)\left(\partial_{\hat{j}} A_{\hat{i}}\right), \label{H}
\end{align}
where $\hat{i},\hat{j}=\{x,y,z\}$. This result can reduce the amount of matter needed to sustain a motion with velocity $v$ compared to the Newtonian version of the same theorem. {Remarkably, the virial theorem can be written in a more suggestive way
\begin{align}
    \langle 2\int \rho v^2 d^3x-\frac{1}{8\pi G} \int \left(E^2+B^2\right) d^3x \rangle=0.
\end{align}
Using the analogy with electromagnetism we see that the second term is the total energy stored in the gravity fields. Therefore we have the balance equation
\begin{align}
    2 \times \text{energy of free dust (kinetic energy)} = \text{energy of gravity}.
\end{align}}

There is another important effect of the gravitomagnetic field: in fact, circular orbits in planes orthogonal to the rotation axis are allowed thanks to the presence of the gravitomagnetic force that balances the Newtonian force in the direction of the rotation axis, which is not possible in purely Newtonian gravity (see e.g \citet{bonnor1977rotating}).\\

The above arguments do not require a specific model for the mass distribution, which is of course important if we want to estimate the order of magnitude of the gravitomagnetic field. The latter, was recently estimated in a paper by \citet{Toth:2021bfs}. In order to evaluate the galactic gravitomagnetic field to estimate its impact on  the rotation curves, the author considers the following gravitomagnetic potential
\beq
\mb A =  \frac{G}{c} \frac{\mb J \times \mb x }{|\mb x|^{3}} \label{eq:solgemAi3}
\eeq 
which corresponds to the case  of a compact source of angular momentum $\mb J$. From this potential it is possible to obtain the gravitomagnetic field  
\beq
\mb B= \frac{G}{c} \left[\frac{3\left(\mb J \cdot \mb x \right)\mb x}{|\mb x|^{5}}- \frac{\mb J}{|\mb x|^{3}} \right] \label{eq:gemfieldspin0}
\eeq
with its dipole-like behaviour. We point out that the gravitomagnetic potential (\ref{eq:solgemAi3}) \textit{is not} a solution of the Poisson equation (\ref{eq:poisson02}) \textit{within} the mass distribution, but \textit{in vacuum.}  Consequently, it is hard to accept that the expression (\ref{eq:gemfieldspin0}) can be used to estimate the galactic gravitomagnetic field. In addition, in doing so,  it is assumed that the gravitomagnetic field at a given location is determined only by the internal mass distribution: the underlying idea is that  the gravitational field is determined by the internal mass distribution only, in analogy with what happens in Newtonian gravity under suitable symmetry hypotheses. Actually, things are more complicated when we are dealing with gravitomagnetic fields in GR: for instance the gravitomagnetic field nearby the center of a rotating mass ring (see \citet{ruggiero2016gravitomagnetic}) is not null, but it is given by
\beq
\mb B=\frac{2G}{cR^{3}}\mb J \label{eq:GM1}
\eeq
where $R$ is the radius of the ring and $\mb J$ its angular momentum. Or, if we consider a uniformly rotating hollow homogeneous sphere, the gravitomagnetic field (see \citet{ciufolini2003gravitomagnetic}) is
\beq
\mb B=\frac{4GM}{3cR} \bm \Omega, \label{eq:GM2}
\eeq
where $M$ is the mass of the sphere, $R$ its radius, and $\bm \Omega$ its angular velocity. Notice that in the latter case, \textit{the corresponding gravitational field is null:} this shows that it is not generally true that gravitomagnetic fields are always smaller than the Newtonian ones.\\ 
Accordingly, we suggest that the estimate of the galactic gravitomagnetic field obtained by \citet{Toth:2021bfs} is based on an oversimplified model and, hence, cannot be used as an argument \textit{against} the impact of GR on galactic rotation curves.

\section{The general relativistic axisymmetric stationary system coupled to dust \label{GRS}}

Following \citet{Astesiano:2021ren}, to describe a single disc galaxy we  consider neutral, stationary and  axisymmetric dust coupled to Einstein's equations. Using cylindrical coordinates $\left(ct,r,\phi,z \right)$ with space-time signature $(-1,1,1,1)$\footnote{For the sake of simplicity, in this Section we use units such that $c=1$.},  matter is assumed to flow along the Killing vectors $\partial_t$ and $\partial_\phi$ and, here and henceforth, functional dependence on the coordinates $(r,z)$ - which are not associated to Killing vectors - is allowed only. If $\rho$ denotes the  matter density, the energy momentum tensor is given by
\begin{align}
    T^{\mu\nu} (r,z)= \rho(r,z) u^{\mu}(r,z) u^{\nu}(r,z), \quad u^{\mu}(r,z)= \frac{1}{\sqrt{-H(r,z)}} \left(1,0,0,\Omega(r,z) \right), \label{EMT}
\end{align}
where $\Omega(r,z)= \frac{d\phi}{dt}= \frac{u^{\phi}}{u^{t}}$. As shown by \citet{stephani_kramer_maccallum_hoenselaers_herlt_2003}, the solution of Einstein's equations is completely determined by the choice of a negative function $H(\eta)$, on which the physical properties depend. Afterwards, it is possible to obtain an  auxiliary function $\mathcal{F}(\eta)$ using\footnote{Here and henceforth, we use the following notation: for any function of one argument, like $H(\eta)$, with a prime we mean the derivative with respect to its argument; in addition, we use a comma to indicate partial derivative with respect to  a given coordinate.}
\begin{equation}
\label{def v}
	\mathcal{F}=2\eta+r^2\int\frac{H'}{H}\frac{d\eta}{\eta}-\int\frac{H'}{H}\eta d\eta.
\end{equation}%
The remaining equations of motion are 
\begin{gather}
\label{harm}
	\mathcal{F}_{,rr}-\frac{1}{r}\mathcal{F}_{,r}+\mathcal{F}_{,zz}=0, \\
	\Omega=\frac{1}{2}\int H'\frac{d\eta}{\eta} \label{eta cond}. 
\end{gather}
After choosing $H(\eta)$ and the solution of Eq. (\ref{harm}), we can calculate the metric components as
\begin{align}
	g_{tt}&=\frac{(H-\eta\Omega)^2-r^2\Omega^2}{H}, \cr
	g_{t\phi}&=\frac{\eta^2-r^2}{(-H)}\Omega+\eta, \cr
	g_{\phi\phi}&=\frac{r^2-\eta^2}{(-H)} \label{components}.
\end{align}%
In addition, the remaining metric components
\begin{align}
g_{zz}=g_{rr}=:e^{\Psi} 
\end{align}
are determined by the following equations
\begin{align}
    \Psi_{,r} =& \frac{1}{2r} \left[ (g_{tt})_{,r} (g_{\phi\phi})_{,r}-(g_{tt})_{,z} (g_{\phi\phi})_{,z} - ((g_{t\phi})_{,r} )^2+((g_{t\phi})_{,z} )^2 \right]     \label{mur},\\
    \Psi_{,z}= & \frac{1}{2r} \left[ (g_{tt})_{,z} (g_{\phi\phi})_{,r} + (g_{tt})_{,r} (g_{\phi\phi})_{,z} -  2 (g_{t\phi})_{,r} (g_{t\phi})_{,z}  \right] \label{muz}.
\end{align}
Eventually, the matter density is given by
\begin{equation}
\label{rho}
	8\pi G\rho=\frac{\eta^2 r^{-2}(2-\eta l)^2-r^2l^2}{4g_{rr}}\frac{\eta_{,r}^2+\eta_{,z}^2}{\eta^2},
\end{equation}
where  $\displaystyle l= \frac{H'}{H}$.\\
Since it will be useful in what follows, we notice that Eq. (\ref{eta cond}) corresponds to the two following conditions, with a little abuse of notation:
\begin{align}
    H_{,r}- 2 \eta \Omega_{,r}=0, \quad  H_{,z}- 2 \eta \Omega_{,z}=0. \label{eta cond2}
\end{align}

\textcolor{black}{We suppose that a galaxy has a finite extension: as a consequence,  flatness at space infinity $r,z \rightarrow \infty$ is expected, which means that, in this limit, the metric reduces to the Minkowski one:
\begin{align}
    g_{tt}=-1,\quad g_{t\phi}=0, \quad g_{\phi\phi}=r^2,\quad g_{rr}=g_{zz}=1.
\end{align}}

\subsection{The projection along the word-lines of the ZAMO} \label{ZAMOSS}

The space-time metric which describes our model of galaxy is stationary and axisymmetric; in this case, care must be paid in choosing suitable observers. In fact, it is known that the use of static observers at rest as seen from infinity is not a good choice, because these observers are not defined by local properties of space-time and, in addition, they cannot exist in some regions  (see e.g. \citet{1970ApJ...162...71B}). So, rather than referring quantities to a coordinate frame,  it is better to use an orthonormal tetrad carried by the so-called ``locally nonrotating observers'' or ``zero angular momentum observers'' (ZAMO), since their angular momentum vanishes. It turns out that these observers are natural candidates to analyse physical processes in the simplest way, since their motion compensates, as much as possible, the dragging effect due to the angular momentum of the source \cite{1972ApJ...178..347B,thorne1986black}. Accordingly, we choose  the ZAMO  to describe the dust motion. As we said,  these observers are non-rotating, in the sense that they are orthogonal to the constant time spacelike hypersurfaces $\Sigma_t$ and define a field of one-forms
\begin{gather}
    n= \frac{r}{\sqrt{g_{\phi\phi}}} dt. \label{E0ZAMO}
\end{gather}
The full orthonormal frame they carry with themselves can be constructed choosing
 \textcolor{black}{
\begin{gather}
     e^{(t)}= n, \qquad   e^{(\phi)}=\sqrt{g_{\phi\phi}} (d\phi-\chi dt),\qquad  e^{(r)}= e^{\Psi/2}\, dr,\qquad  e^{(z)}= e^{\Psi/2}\, dz, \label{VierbeinZamo}
\end{gather}}
where, for simplicity, we defined
\begin{gather}
\chi \equiv - \frac{g_{t\phi}}{g_{\phi\phi}}=  \frac{H \eta}{(r^2-\eta^2)}+\Omega.    \label{eq:defchi}
\end{gather}
\textcolor{black}{Actually, $\chi$ is the angular velocity of the ZAMO frame as seen from an asymptotic inertial observer   at infinity  \cite{1972ApJ...178..347B}.
Notice  that, thanks to the definition (\ref{eq:defchi}), the ZAMO consistently satisfies the requirement of zero angular momentum: $\displaystyle g_{\phi\phi}\, \chi+g_{\phi t}=0.$}

The relevant elements of the dual basis are found to be
\textcolor{black}{
\begin{gather}
      e_{(t)}=\frac{r}{\sqrt{g_{\phi\phi}}} g^{tt} \left( \partial_t+ \frac{g^{t\phi}}{g^{tt}} \partial_\phi \right)=  \frac{\sqrt{g_{\phi\phi}}}{r} \left( \partial_t+ \chi \partial_\phi \right),  \\
     e_{(\phi)} =\frac{1}{\sqrt{g_{\phi\phi}}} \partial_\phi.
\end{gather}}
Using these definitions,  the four velocity of the dust (\ref{EMT}) can be rewritten as

\textcolor{black}{
\begin{align}
    u= \frac{r}{\sqrt{g_{\phi\phi}}} \frac{1}{\sqrt{-H}} \left(e_{(t)}+ \frac{{g_{\phi\phi}}}{r} (\Omega-\chi) e_{(\phi)}\right),     \label{FVD}
\end{align}}

\textcolor{black}{On the other hand, the three-velocity  of the dust measured by the ZAMO is
\beq
v^{(a)}=\frac{u^{\mu}e_{\mu}^{(a)}}{u^{\mu}e_{\mu}^{(t)}}.
\eeq}
Using the expressions (\ref{FVD}) of the four velocity and the metric components (\ref{components}), and  taking into account the orthonormal features of the tetrad, we obtain
\beq
v^{(\phi)}=\frac{u^{\mu}e_{\mu}^{(\phi)}}{u^{\mu}e_{\mu}^{(t)}}=\frac{\eta}{r} \doteq v.  \label{eta}
\eeq
As a consequence, we can use $v$ to give a physical meaning to the mathematical function $\eta$: as we are going to show, this function is simply related to the angular momentum per unit mass of a dust element.

In addition we may define
\textcolor{black}{
\begin{align}
    u^{\mu}e_{\mu}^{(t)}= \frac{1}{\sqrt{1-v^2}} \doteq  \gamma. \label{eq:defgamma}
\end{align}}
Exploiting the above definitions, it is possible to write the four-velocity vector of the dust (\ref{FVD}) in the very simple form
\beq
u=\gamma\left(e_{(t)}+v\, e_{(\phi)} \right). \label{SPLIT1}
\eeq
Since $\displaystyle \frac{r}{\sqrt{g_{\phi\phi}}}=\sqrt{-H}\gamma$, we may write the elements of the basis tetrad in the form
\beq
e^{(t)}=\sqrt{-H}\gamma dt, \quad e^{(\phi)}=\frac{r}{\sqrt{-H}\gamma}(d\phi-\chi dt), \label{eq:redefbasis}
\eeq
so that the metric turns out to be
\textcolor{black}{
\begin{align}
    ds^2&= H\gamma^{2}{} dt^2 - r^2 \frac{1}{H\gamma^{2}}\left(d\phi-\chi dt \right)^2+ e^{\Psi}\left(dr^2+dz^2\right). \label{IR} 
\end{align}}
Notice that from Eq. (\ref{eq:defchi}) we obtain
\textcolor{black}{
\beq
\chi= \frac{v}{r} H\gamma^{2}+ \Omega \label{Omega2}
\eeq}
which can be written as
\begin{align}
    r \Omega = r \chi  -v \gamma^2\, H.  \label{Chi}
\end{align}
This can be seen as a generalisation of the usual relation $v = r \Omega$ of  Newtonian mechanics, which is restored in the limit where the effects of $\chi$ are negligible and $-\gamma^2\, H  \rightarrow 1$.  We notice  that $r\Omega$ represents the \textit{coordinate velocity}, i.e. the velocity of the dust as measured by inertial observers at rest at infinity. 

In addition, we may write the energy momentum tensor (\ref{EMT}) in the form
\begin{align}
    T= \rho u \otimes u= \rho \gamma^2 \left(e_{(t)}+v\, e_{(\phi)}\right) \otimes \left(e_{(t)}+v\,e_{(\phi)}\right),
\end{align}
from which it is now easy to read the corresponding tetrad components
\begin{align}
    T^{(t)(t)}= \gamma^2\rho, \quad T^{(t)(\phi)}= \gamma^2 \rho v, \quad T^{(\phi)(\phi)}= \gamma^2 \rho v^{2}.
\end{align}
The Killing vectors $\partial_t$ and $\partial_\phi$ define associated conserved quantities along the flow of the dust
\begin{align}
   \mathcal{E}:=- u^\mu (\partial_t)_\mu&= \frac{1}{\sqrt{-H}} \left[\gamma^2 (-H)+ r v \chi\right]= \frac{1}{\sqrt{-H}} \left[\gamma^2 (-H)+ \eta \chi\right] \label{Ene} \\
   \mathcal{M}:=u^\mu (\partial_\phi)_\mu&= \frac{1}{\sqrt{-H}} r v=\frac{1}{\sqrt{-H}} \eta \label{Mom}
\end{align}
The first one $ \mathcal{E}$ is the energy (per unit mass), while the second one $\mathcal{M}$ is the angular momentum (per unit mass) which gives a physical interpretation to $\eta$. \\

\subsection{Light frequency shift measured from an inertial asymptotic observer } \label{ssec:redshift}

After having analysed, using the ZAMO, the features of the metric which constitutes our model of a galaxy,  we focus on what can be measured by asymptotically inertial observers to investigate galactic dynamics: namely,   the frequency shift of a photon emitted by a particle of dust. 
To this end, the following hypotheses are made: (i) the emitters are supposed to move in (stable) circular geodesics; (ii) after the emission, photons  propagate along null geodesics, so that any possible refraction effects are neglected.
In our case, the frequency shift is the sum of the gravitational and Doppler contributions: in fact, the photon is emitted from an object  moving in a gravitational field. We assume that a photon is emitted with proper frequency $\nu_e$ by a dust particle, while  $\nu_d$ is the frequency measured by the detector. The frequency shift is measured by the redshift factor $\tilde{z}$ (negative for a blueshift), which is in general defined by
\begin{equation}
\label{def redshift}
	1+\tilde{z}:=\nu_e/\nu_d.
\end{equation}
The proper frequencies are (see e.g. \citet{Ruggiero:2006cq})
\begin{equation}
\label{energie}
	\nu_{e,d}:=-U^{\mu}_{e,d}k_{\mu},
\end{equation}%
where $U^{\mu}_d$ is the 4-velocity of the detector, $U^{\mu}_e$ is the 4-velocity of the source, and $k_{\mu}$ is the 4-momentum of the photon at the respective locations.\\
Under the assumption that gravitational effects could be neglected when measuring the frequency shift of  light coming from an external galaxy, the measured redshift $\tilde{z}$ would just be comparable to the pure kinematic Doppler effect in a Minkowski space-time, corresponding to a ``special-relativistic''  velocity $\mb{v}_{SR}$ of the source, i.e., 
\begin{equation}
	1+z=\frac{1+v_{SR}^{||}}{\sqrt{1-v_{SR}^2}}.
\end{equation}%
The vector $\mb{v}_{SR}$ is assumed to be directed along $\partial_\phi$ and we call its projection along the line of sight  $v_{SR}\sin\theta$. \\
The general relativistic description leads to a more general and interesting result. Let us now compare the accepted special relativistic description (SR), which accounts only for the kinematic Doppler shift, to the general relativistic description (GR), which instead includes the gravitational shift effect. The expressions for the redshift for the model that we are considering were calculated by \citet{Astesiano:2022tbh}, and read:
\begin{equation}
1+z=
\begin{cases}
    \text{SR}\quad  \frac{1+v_{SR}\sin\theta}{\sqrt{1-v_{SR}^2}}, \\
    \text{GR}\quad \frac{1}{\sqrt{-H}}\left[1+\frac{r\Omega\sin\theta}{\sqrt{(\gamma^2H)^2-(\gamma^2Hv+r\Omega)^2\cos^2\theta}-(\gamma^2Hv+r\Omega)\sin\theta}\right], \label{GRRedshift}
\end{cases}
\end{equation}
where $\frac{\pi}{2}-\theta$ is  the angle between $\partial_\phi$ and the emitted photon.\\
We see that the overall effect depend both on the kinematical   effects ($v$) and gravitational  ones ($\gamma^2 (-H) , \chi$). The degree of freedom given by the non diagonal term $\chi$ affects the result, although it is not explicit in the formula. For example we remember that $r\Omega \neq v$ when $\chi$ is not negligible (see Eq. (\ref{Chi})).\\
As a particular application of this result,  if the galaxy is seen edge-on ($\theta=\pm \frac{\pi}{2}$) the resulting redshift turns out to be
\begin{align}
    1+\tilde{z}^{||}= \frac{(\gamma\sqrt{-H})}{\gamma^2(-H) \mp r \chi} \frac{\left(1\pm v\right)}{\sqrt{1-v^2}}, \label{cut2}
\end{align}
where the upper and lower sign refers to backward and forward emission, respectively. 
If the galaxy is seen face-on ($\theta=0$) we obtain the following result:
\begin{align}
    1+\tilde{z}^{\perp}=\frac{1}{\sqrt{-H}}. \label{Transverse}
\end{align}
The same result is obtained if we are observing a disc galaxy tilted with a certain angle with respect to the line of sight and we perform the measurement on the minor axis (see, again \citet{Astesiano:2022tbh}).

\section{Some relevant low energy limits} \label{sec:limits}

Here, we want to further investigate the properties of the general solution studied so far. We expect that  a galaxy is a low energy system: as we said, it is reasonable to suppose that  far from the galactic center the gravitational field is weak  and stars are not moving at relativistic speeds. Accordingly, we will expand the coefficients of the  metric (\ref{IR}) in negative powers\footnote{Hence, physical units are restored throughout this Section.} of $c$ and make a comparison with known limits of the solutions of Einstein's equations: this will help us to obtain a physical interpretation for the functions 
$\gamma^2(-H)$ and $\chi$, that are respectively related to the Newtonian potential and to the gravitomagnetic potential.

The first relevant terms in the low energy expansion of these functions that we consider are given by
\begin{align}
    \gamma^2(r,z) (-H)(r,z)=&\, \mathcal{A}(r,z)_{(0)} c^{0}+\mathcal{A}(r,z)_{(-2)} c^{-2}+ O(c^{-4}), \label{NP}\\
    \chi(r,z)=&\, \chi(r,z)_{(-1)} {c^{-1}}+\chi(r,z)_{(-2)} {c^{-2}}+O(c^{-3}). \label{GP}
\end{align}
If, in the above expressions, we consider only the $\mathcal{A}_{(0)}$ term,  we get the special relativistic limit, as it will be shown in Section \ref{SR}; on the other, if we take into account also the $\mathcal{A}_{(-2)}$ term,  we are led to the usual Newtonian limit, as discussed in Section \ref{Newtonian}. Further information are obtained by considering the effects of the non diagonal terms,  thus allowing $\chi_{(-1)}$ to be different from 0; in particular, in Section \ref{GEM} we propose a more general limit where the off-diagonal terms are of the same order of the Newtonian effects, which we call the strong gravitomagnetism (SGM) case: this model provides a simple explanation of the flat velocity profile of disc galaxies. Moreover, in this general relativistic context it is possible to discuss the mass density needed to produce a flat velocity profile and  compare it to what is obtained in Newtonian gravity. In Appendix \ref{APP} we show that the term $\chi(r,z)_{(-2)}$ gives rise to the usual gravitomagnetism, discussed in Section \ref{Gravitomagn}.
Eventually, in Section \ref{Rigid}, we will consider as a particular case the rigidly rotating model $\Omega=$ constant, which was used as a model for our galaxy in previous publications \cite{Balasin:2006cg,crosta2020testing}.\\
In what follows, we will discuss the impact of these limits on the measured redshift. According to the general  expansion given by eq. (\ref{NP}) and (\ref{GP}), the form of the redshift (\ref{GRRedshift}) at first order in $c$ becomes
\begin{equation}
1+z=
\begin{cases}
    \text{SR}\quad  1+\frac{v_{SR}}{c}\sin\theta+ O(c^{-2}), \\
    \text{GR}\quad  1+\frac{r\Omega\sin\theta}{c}+O(c^{-2}),  \label{GRR}
\end{cases}
\end{equation}
where $\Omega=\frac{v}{r}+\chi$ at the leading order. Thanks to the above expressions it is  manifest that, when the off diagonal term $\chi$ is negligible, the general relativistic result coincides with the special relativistic one at the first order, with the identification $r\Omega= v= v_{SR}$.

\subsection{The special relativistic limit } \label{SR}

Setting $\mathcal{A}(r,z)_{(0)}=1$ and neglecting the other contributions from the expansions (\ref{NP}) and (\ref{GP}) we obtain
\begin{align}
        \gamma^2 (-H)= 1, \quad \chi= 0.
\end{align}
Hence, taking also into account Eqs. (\ref{mur}),(\ref{muz}) we get the special relativistic limit, i.e.:
\begin{align}
    ds^2&= - c^2 dt^2 + r^2 d\phi^2+ dr^2+dz^2. \label{eq:flat1}
\end{align}
Of course this is a limiting case, since there is no matter as source of the gravitational fields ($\rho=0$, see  Eq. (\ref{rho})) and then the four vector of the dust $u$ can be seen as referring to free particles in the metric (\ref{eq:flat1}).\\
In this limit the parallel and transverse redshifts in e Eqs. (\ref{cut2})-(\ref{Transverse})  give, respectively:
\begin{align}
    1+\tilde{z}^{||}=&  \frac{\left(1\pm \frac{v}{c}\right)}{\sqrt{1-\frac{v^2}{c^2}}}= \pm \frac{v}{c}+ \frac{v^2}{2c^2}+O(c^{-3}),  \\
    \quad 1+\tilde{z}^{\perp}=& \frac{1}{\sqrt{1-\frac{v^2}{c^2}}}=\frac{v^2}{2c^2}+O(c^{-3}),
\end{align}
as expected from the usual relations for the Doppler effect. Therefore, we have a simple relation between the two redshifts
\begin{align}
    \tilde{z}^{||}= \pm \frac{v}{c}+ \tilde{z}^{\perp}+ O(c^{-3}). \label{RR}
\end{align}

\subsection{The Newtonian limit } \label{Newtonian}

In this Section, we will show how to obtain the Newtonian limit up to order $c^{-2}$, therefore neglecting post-Newtonian corrections. Anyhow, the post-Newtonian limit can be obtained just adding the appropriate terms in the expansion (\ref{NP}).\\
The Newtonian limit is obtained by taking
\begin{align}
    \gamma^2 (-H)= 1- \frac{2U(r)}{c^2}+ O(c^{-4}), \quad \chi= 0 ,
\end{align}
where $U(r)$ is the Newtonian potential; consequently,  up to order $c^{-2}$, we may write
\begin{align}
    -H(r)= 1- \frac{2U(r)}{c^2}-\frac{v(r)^2}{c^2}+O(c^{-3}). \label{eq:valacca}
\end{align}
We get a well known result: the Newtonian limit is cylindrically symmetric.  Equation (\ref{Omega2}) gives 
\begin{align}
    \Omega \, r = v+ O(c^{-3}).  \label{eq:omegacordv}
\end{align}
In this case, the coordinate velocity $r\Omega$ equals the velocity measured by the ZAMO: this is not a surprise, since in this case the ZAMO  are the same as the observers at rest at infinity.  Since we already know the function $\Omega$, we can use it to impose Eq. (\ref{eta cond}) (or equivalently Eq. (\ref{eta cond2})), to obtain 
\begin{align}
    \partial_r U= - \frac{v^2}{r}, 
\end{align}
from which we directly read the condition for circular orbits. 

In this context, we can check what happens to the auxiliary function $\mathcal{F}$, defined by Eq. (\ref{def v}). To this end, we calculate the integrals
\begin{align}
    c^2 r^2 \int \frac{H'}{H} \frac{d\eta}{\eta}&=-c^2 r^2 \int H' \frac{d\eta}{\eta}+ O(c^{-2})= -2 r^2 \Omega+ O(c^{-2})= -2 r v+ O(c^{-2}), \\
    -\int \frac{H'}{H} \eta d\eta &= -\int \frac{\partial_r H }{H} \frac{vr}{c} dr =0 + O(c^{-2}).
\end{align}
Therefore $\mathcal{F}=0$ up to order $c^{-2}$ and automatically solves Eq. (\ref{harm}).\\ In this case, the gravitomagnetic effects, related to the off-diagonal terms of the metric  $g_{t\phi}$ are suppressed, since
\begin{align}
    g_{t\phi}=0+O(c^{-3}), \quad g_{tt}= -c^2  \left(1+ \frac{2U}{c^2} \right)+O(c^{-2}),\quad g_{\phi\phi}= r^2 \left(1+\frac{2U}{c^2}\right)+ O(c^{-4}).
\end{align}
Eventually, these results allow us to obtain from Eq. (\ref{mur}) the last unknown element of the metric
\begin{align}
    e^{\Psi}= e^{2\frac{U}{c^2}+ O(c^{-4})}= 1+ \frac{2U}{c^2}+ O(c^{-4}).
\end{align}
With all the ingredients, we can write down the metric $(\ref{IR})$ 
\begin{align}
    ds^2=c^2\left(-1+ \frac{2U}{c^2}\right) dt^2+ \left(1+ \frac{2U}{c^2}\right) \left(r^2 d\phi^2+ dr^2+dz^2 \right)+ O(c^{-3}),
\end{align}
while the  density (\ref{rho}) is
\begin{align}
    \frac{8 \pi G}{c^2} \rho= 4 \frac{v}{c^2 r} \partial_r v+  O(c^{-4}) \label{Nd},
\end{align}
in agreement with Eq. (\ref{eq:rhoomega}).  The energy and the angular momentum are
\begin{align}
     \frac{\mathcal{E}}{c^2}=&1+\frac{v^2}{2c^2}- \frac{U}{c^2}+O(c^{-4}),  \\
     \mathcal{M}=& v r+ O(c^{-3})= \Omega^2 r+O(c^{-3}) .
\end{align}
The above results can be used to get further insight into the application of a Newtonian approach to the description of galactic dynamics. In fact, a flat velocity profile in the Newtonian regime would lead to $v=$ constant\footnote{Remember that in this case $v$ equals the coordinate velocity (see e.g. Eq. (\ref{eq:omegacordv})).} as can be seen from Eq. (\ref{cut2}). This cannot be achieved since imposing the constraint $v=$constant implies $\rho=0$ in Eq. $(\ref{Nd})$.\\

The parallel and transverse redshifts in Eqs. (\ref{cut2})-(\ref{Transverse})  are, respectively:
\begin{align}
    1+\tilde{z}^{||}( \pm\pi/2, r, z)=  \frac{\left(1\pm \frac{v}{c}\right)}{\sqrt{1-\frac{v^2}{c^2}- \frac{2U}{c^2}}}, \quad 1+\tilde{z}^{\perp}( \pm\pi/2, r, z) = \frac{1}{\sqrt{1-\frac{v^2}{c^2}- \frac{2U}{c^2}}}.
\end{align}
Therefore, we have the same relation as in the special relativistic case Eq. (\ref{RR}) between the two redshifts
\begin{align}
    \tilde{z}^{||}= \pm \frac{v}{c}+ \tilde{z}^{\perp}+ O(c^{-3}), \label{RRN}
\end{align}
As a consequence,  we may neglect  the frequency shift due to the gravitational field in the Newtonian approximation. As it is clear from the above relations, the Doppler effect is of order $c^{-1}$ while the gravitational shift is of order $c^{-2}$.

\subsection{The ``strong gravitomagnetic" (SGM) limit and a proposed model for disc galaxies} \label{GEM}

The presence of off-diagonal or \textit{gravitomagnetic} terms in the metric leads to the \textit{dragging of inertial frames}: a gyroscope, which defines the orientation of a local inertial frame, rotates relative to observers at rest at infinity, because the gravitational field of the source drags the space-time around it. This effect can be explained as the action of the gravitomagnetic field of the source on the gyroscope spin (see e.g. \citet{MTW,ciufolini1995gravitation,bosi2011measuring}).
The analogy can be done with a solid sphere, rotating in a  viscous fluid: because of its rotation, the fluid is dragged along with the sphere. 
In order to take into account this  effect, in the  low energy expansion (\ref{NP})-(\ref{GP}), we consider a dragging term in addition to the usual Newtonian potential; in other words, we take the expansion defined by
\begin{align}
    \gamma^2 (-H)= 1- \frac{2U(r,z)}{c^2}+ O(c^{-3}), \quad \chi= \frac{a(r,z)}{r^2}+O(c^{-1}), \label{eq:termdrag}
\end{align}
and call it the "strong gravitomagnetic" limit (SGM).  Accordingly, in this case Eq. (\ref{Omega2}) gives
\begin{align}
    \Omega(r,z)= \frac{a(r,z)}{r^2} + \frac{v(r,z)}{r}. \label{eq:composiz}
\end{align}
It is interesting to rephrase Eq. (\ref{eq:composiz}) in the form
\beq
r \Omega = r \chi+v   \label{eq:addition}
\eeq
which can be interpreted as a classical velocity-addition relation: the velocity $r \Omega$, measured by inertial observers,  is the sum of the velocity $v$, measured by the ZAMO,   and  the velocity of the ZAMO with respect to  inertial observers, $r \chi$.
Once again, knowing $\Omega$ we can use it to impose Eq. (\ref{eta cond}) and obtain:
\begin{align}
    \partial_r U+  \frac{v^2}{r}- vr \partial_r \left(\frac{a}{r^2}\right)&=0, \label{ER}\\
    \partial_z U- \frac{v}{r} \partial_z a&=0.\label{EZ}
\end{align}
It is relevant to emphasize the role of the function $a$: we  explicitly see from the above equations that 
when $a=0$, $\partial_z U=0$, hence the Newtonian potential is cylindrical symmetric; this symmetry is broken by the presence of this function.

As for  the auxiliary function $\mathcal{F}$ defined in (\ref{def v}), we calculate the integrals
\begin{align}
   c^2 r^2 \int \frac{H'}{H} \frac{d\eta}{\eta}&=-c^2 r^2 \int H' \frac{d\eta}{\eta}+ O(c^{-4})= -2 r^2 \Omega+ O(c^{-2}), \\
    -\int \frac{H'}{H} \eta d\eta &= -\int \frac{\partial_r H }{H} vr dr =0 + O(c^{-2}),
\end{align}
therefore 
\begin{align}
    \mathcal{F}= -2 \frac{a}{c}.
\end{align}
Using these relations, we obtain the following expressions for the metric elements
\begin{align}
   g_{tt}= -c^2\left(1-\frac{2U}{c^2}-\frac{a^2}{c^2r^2}\right),\quad g_{t\phi}= -a,\quad g_{\phi\phi}=r^2\left(1+ \frac{2U}{c^2}\right), \label{MFD}
\end{align}
with a non-negligible off diagonal term depending on $a$. The equation of motion (\ref{harm}) is written in the  simple form:
\begin{align}
    a_{,rr}- \frac{a_{,r}}{r}+a_{,zz}=0, \label{EOMa}
\end{align}
which is exactly the condition for the integration of Eqs. (\ref{mur}) and (\ref{muz}) which read
\begin{align}
    \Psi_{,r}=\frac{1}{2r} \left[2r \partial_r\left(\frac{2U}{c^2}+ \frac{a^2}{c^2r^2}\right)+ \frac{a^2_{,z}-a^2_{,r}}{c^2} \right]+ O(c^{-4}), \label{psir} \\ \Psi_{,z}=\frac{1}{2r} \left[2r \partial_z\left(\frac{2U}{c^2}+ \frac{a^2}{c^2r^2}\right)-\frac{2}{c^2} a_{,r} a_{,z}\right]+ O(c^{-4}) \label{psiz}.
\end{align}
They can be directly integrated as
\begin{align}
    \Psi= \frac{2U}{c^2}+ \frac{a^2}{c^2r^2}+ \frac{1}{2c^2} \int \frac{a^2_{,z}-a^2_{,r}}{r} dr+ O(c^{-4}),
\end{align}
therefore, the last element of the metric is
\begin{align}
    g_{rr}= g_{zz}= e^\Psi= \left[1+ \frac{2U}{c^2}+ \frac{a^2}{c^2r^2}+\frac{1}{2c^2} \int \frac{a^2_{,z}-a^2_{,r}}{r} dr \right] + O(c^{-4}).
\end{align}
To obtain the energy density (\ref{rho}) for this model, we start evaluating the function $\ell$ and, taking into account Eq. (\ref{eq:valacca}), which holds true also in the dragging limit thanks to the definitions (\ref{eq:termdrag}), we get
\begin{align}
    \ell= \frac{H'}{H}= \frac{(-H)_{,r}}{\eta_{,r}} \frac{1}{(-H)}= \frac{-2 \eta \Omega_{,r}}{c^2\eta_{,r} (-H)}=\frac{-2 \eta \Omega_{,r}}{c^2\eta_{,r}}+ O(c^{-4}).
\end{align}

At order $c^{-2}$ the relevant terms in the energy density are
\begin{align}
    \frac{8\pi G}{c^2} \rho = \left(\frac{v^2}{c^2}-\frac{c^2 r^2}{4}\ell^2\right) \frac{\eta_{,r}^2+\eta_{,z}^2}{\eta^2}. \label{dendrag}
\end{align}
After substituting, we obtain:
\begin{align}
    \frac{8\pi G}{c^2} \rho_D= \frac{1}{c^2} \left[4v \frac{v_{,r}}{r}+2 \left(v-rv_{,r}\right)b_{,r}-r^2 b^2 _{,r}\right] \frac{(v+r v_{,r})^2+r^2 v_{,z}^2}{(v+r v_{,r})^2}, \label{ddz0}
\end{align}
where $b= r^{-2} a$ and we used $\rho_{D}$ to denote the density obtained in this dragging limit. It is interesting to evaluate the difference between this density $\rho_D$ and the Newtonian density $\rho_N$ given in Eq. (\ref{Nd}), for the same value of the velocity $v$ in both models. Since the Newtonian limit is cilindrically symmetric we focus on the equatorial plane $z=0$ (where for symmetry $\partial_z v=0$)
\begin{align}
   \frac{8\pi G}{c^2} \delta \rho\equiv \frac{8\pi G}{c^2}\rho_D(z=0)-\frac{8\pi G}{c^2}\rho_N= \frac{b_{,r}}{c^2} \left[ 2(v-rv_{,r})-r^2 b_{,r}\right].  \label{deltadens}
\end{align}
It is clear that the presence of a non-diagonal dragging term $a$ greatly affects the density required to sustain the motion. As we will show in Appendix (\ref{APP}), the first term is present also in the gravitomagnetic limit, while the pure negative term $-r^2 b^2_{,r}$ is a peculiar feature of the SGM limit: this term can significantly reduce the required mass compared to the Newtonian case, where these effect are not present.

Hence, the analysis of exact solutions leads to the same conclusion obtained in Section \ref{Gravitomagn} in the weak-field limit: non-diagonal terms in the space-time metric can lead to a re-evaluation of the weight of dark matter in galaxies.
\\
For completeness, we calculate the values of the energy (\ref{Ene}) and  angular momentum (\ref{Mom}) in this limit:
\begin{align}
     \frac{\mathcal{E}}{c^2}=&1+\frac{v^2}{2c^2}- \frac{U}{c^2}+ \frac{av}{rc^2}+O(c^{-4}), \label{Energia} \\
     \mathcal{M}=& v r+ O(c^{-3}).\label{Momentoangolare}
\end{align}

\section{Redshift analysis of the rotation curves } \label{eq:analysis} 

Here we will apply the results obtained so far  to show that it could be possible to propose a SGM model for a disc galaxy which agrees with the current observations of a flat velocity profile. In doing so, we will point out the effect of the SGM terms that we discussed in Section \ref{GEM}. Accordingly,  from Eq. (\ref{cut2}) for an edge-on galaxy\footnote{The extension to the case of generic angle $\theta$ is trivial: see Eq. (\ref{GRR}).} we obtain:
\begin{align}
     1+\tilde{z}^{||}= 1 \pm \frac{1}{c}\left(v+ \frac{a}{r}\right)+ O(c^{-2}). \label{Epr}
\end{align}
All  functions depends on $r$ and $z$: however,  to emphasise the physical content, we will restrict the attention to the galactic plane ($z=0$) where, due to the symmetry of the system, we have:
\begin{align}
    \partial_z v|_{z=0}=0, \quad \partial_z a|_{z=0}=0, \quad \partial_z U|_{z=0}=0.
\end{align}
From observations based on redshifts (or blueshifts) measurements, we know that far from the center of the disc galaxy we observe a flat velocity profile:  accordingly, from Eq. (\ref{Epr}), and taking into account the relation (\ref{eq:composiz}) we get
\begin{align}
    \Omega(r,0)= \frac{\alpha}{r}+O(c^{-1}),  \label{Rc}
\end{align}
where $\alpha$ is a constant, defined by $\displaystyle a(r,0)= r \left(\alpha-v(r,0)\right)$.
As already discussed before, a flat velocity profile cannot be obtained in the Newtonian limit because in that case $a=0$; on the other hand, in the SGM limit this observational property can be easily obtained as we have shown above.

From Eqs. (\ref{ER}) and (\ref{EZ}) we get the only condition
\begin{align}
    U(r,0)_{,r}=- v(r,0)\, v(r,0)_{r}-\alpha \frac{v(r,0)}{r},
\end{align}
or their integrated version
\begin{align}
    U(r,0)= - \frac{v(r,0)^{2}}{2}- \alpha \int\frac{v(r,0)}{r} dr.
\end{align}
Then, the energy density in this SGM limit $\rho_{D}(r,0)$ for the flat velocity profile is
\begin{align}
    \frac{8\pi G}{c^2}\rho_{D}(r,0) = \frac{1}{c^2r^2} \left[\left(v(r,0)-r v(r,0)_{,r}\right)^2- \alpha^2\right]+ O(c^{-4}).
\end{align}
Let us now evaluate the transverse redshift given by Eq. (\ref{Transverse}); we obtain: 
\begin{align}
    1+\tilde{z}^{\perp}=\frac{1}{\sqrt{1-\frac{v^2}{c^2}- \frac{2U}{c^2}}},
\end{align}
exactly as in the Newtonian limit, see Eq. (\ref{RRN}). Therefore, the only knowledge of the transverse redshift is not enough to discriminate between the Newtonian and the SGM limits. At order $c^{-2}$ we can write
\begin{align}
    \tilde{z}^{\perp}= \frac{U}{c^2}+ \frac{v^2}{2c^2}+O(c^{-4}),\quad \tilde{z}^{||}= \pm \frac{1}{c} \left(v+\frac{a}{r}\right)+O(c^{-3}).
\end{align}
If we are able to measure both $\tilde{z}^{\perp}$ and $\tilde{z}^{||}$, we can fix both $U$ and $a$ 
\begin{align}
    U= c^2 \tilde{z}^{\perp}- \frac{v^2}{2}, \quad \frac{a}{r}= c\tilde{z}^{||}- \frac{v}{c},
\end{align}
leaving $v$ as the only unknown function which, in turn, can be found using the equations of motion (\ref{ER}) and (\ref{EZ})
\begin{align}
    \partial_z(\tilde{z}^{\perp})=& \pm \frac{v}{c} \partial_z(\tilde{z}^{||}), \\
    \partial_r(\tilde{z}^{\perp})=& \pm \frac{v}{c} \partial_r(\tilde{z}^{||})\mp \frac{v}{cr} \tilde{z}^{||}.
\end{align}

\section{The rigidly rotating case \label{Rigid}}

We discuss in  some detail the case of constant angular velocity $\Omega(r,z)=\Omega_0$, considered by \citet{Balasin:2006cg} and \citet{crosta2020testing} as a model for our galaxy. The first feature is that in this rigidly rotating solution the dust fills the entire space-time, since the equations of motion for constant angular velocity of the matter $\Omega$ are not consistent with $\rho=0$. This rigidly rotating case can be seen as a very particular case of the SGM limit considered before.\\
Even though in this case the function $H(r,z)$ is fixed to $H=-1$, the presence of the off-diagonal term $\chi$ allows for a non trivial profile for $v$, see Eq. (\ref{Chi})
\begin{align}
    \chi= \Omega_0 - \frac{v}{r} \frac 1{1-\frac{v^2}{c^2}} = \Omega_0 - \frac{v}{r} + O(c^2).
\end{align}
We note that rigid rotation does not implies $r^{-1}v$ constant because, $v$ is the velocity measured by the ZAMO and $\Omega_{0}$ is the angular velocity measured from an asymptotic inertial observer. Using the expressions for $H$ and $\chi$ given above, the redshift given by Eq. (\ref{cut2}) takes the simple form
\begin{align}
    1+\tilde{z}(\pm\pi/2, r, z)=1 \pm r \frac{\Omega_0}{c}+ O (c^{-2}); \label{LR}
\end{align}
this linear behaviour is expected for the redshift in from rigidly rotating system. Such a behaviour is not reproduced generally in disc galaxies, except for the inner regions. 
\\
For  constant angular velocity $\Omega_0$, we can perform a rigid rotation of the coordinates
\begin{align}
    \phi'= \phi- \frac{\Omega_0}{c} ct
\end{align}
to rewrite the four velocity of the dust given by (\ref{EMT}) as
\begin{align}
    u= \partial_{t'},
\end{align}
the rotation is equivalent to impose $\Omega_0=0$ everywhere in the coordinates $(t,r,z,\phi)$. Then, restoring the notation without the $'$, the metric (\ref{components}) is now given by
\begin{gather}
    ds^2 = - \left(cdt-\frac{\eta}{c} d\phi\right)^2+ r^2 d\phi^2+e^{\Psi}\left(dr^2+dz^2\right),  \label{RRD}\\
    \eta_{,rr}+\eta_{,zz}-\frac{\eta_{,r}}{r}=0,\quad \Psi_{,r}= \frac{(\eta_{,z})^2-(\eta_{,r})^2}{2r}, \quad \Psi_{,z}=- \frac{\eta_{,r}\eta_{,z}}{r}.  \label{RRD1}
\end{gather}

Notice that the metric in the form (\ref{RRD1}) is exactly the one used by \citet{Balasin:2006cg,Cooperstock:2006dt} and, subsequently, by \citet{crosta2020testing}.

The energy density $\rho(r,z)$ given by (\ref{rho}) boils down to
\begin{align}
    \frac{8 \pi G}{c^2} \rho= \frac{e^{-\Psi}}{c^2} \frac{(\eta_{,r})^2+(\eta_{,z})^2}{r^2}.  \label{density}
\end{align}
The rigidly rotating dust metric in Eq. (\ref{RRD}) in the ZAMO frame reads:
\begin{align}
   ds^2=- \frac{c^2}{1-\frac{v^2}{c^2}} dt^2 + r^2(1-\frac{v^2}{c^2}) \left(d\phi-\chi dt\right)^2+ (e^2)^2+(e^3)^2.  \label{ZAMO2}
\end{align}
As shown in \citet{Astesiano:2022gph}, we can get physical insight into this solution: namely, the gravitational potential $U$ due to the presence of the dust exactly balance the gravitational potential $U_C$ of the non inertial force determined by the rotation of the reference frame.

Using the formalism introduced in Section (\ref{ZAMOSS}), the potential is
\begin{align}
   U= \gamma \sqrt{(-H)}=\frac{1}{\sqrt{1-\frac{v^2}{c^2}}}= 1+ \frac{1}{2} \frac{v^2}{c^2}+O(c^{-3}),
\end{align}
therefore $U=\frac{v^2}{2c^2}$. The balance equation is
\begin{align}
    U+ U_C=0, \label{Balance}
\end{align}
where $\displaystyle U_C= -\frac{1}{2}\frac{v^2}{c^2}$. This equality is the reason why $g_{tt}=-c^2$  \textcolor{black}{in (\ref{RRD});} as a consequence,  if a light signal is emitted from a generic element of the galaxy and it is received from another element, the measured redshift (or blueshift) is 0: the frequency of light does not change \cite{Astesiano:2022gph}.

Eventually,  the conserved quantities given in Eqs. (\ref{Ene}) and (\ref{Mom}) read:
\begin{align}
    \mathcal{E}=c^2,\quad \mathcal{M}= r v,
\end{align}
Clearly, the reason why the energy $\mathcal{E}$ is equal to $c^2$ is again the balance Eq. (\ref{Balance}).\\

\citet{Balasin:2006cg} note the discrepancy between the density in the rigidly rotating case and in the Newtonian case. This is not due to some misterious effects but because these two limits (i.e. the rigid and the Newtonian one) are different limiting cases of the same class of exact solutions and they coincide only in a single point, that is where $r^{-1}v $ is constant and equal to $\Omega_0$. In fact in this point $\chi=O(c^{-2})$ and the densities $(\ref{density})$ and $(\ref{Nd})$ coincide. 

\section{Final remarks and perspectives}

Explaining the observed flat velocity profile in disc galaxies is one of the most challenging problems in current astrophysics. Motivated by various suggestions in the literature, which contributed to focus on the role of General Relativity in this context, we analysed the impact of post-Newtonian corrections on the description of rotation curves. 

To begin with, under the hypothesis that the gravitational field of a galaxy can be considered sufficiently week in its outer regions, we started from a well known low energy limit and
used the gravitomagnetic analogy to show that  the coupling between the fluid vorticity and the gravitomagnetic field  leads to a local relation between density and  velocity which is different from the Newtonian case: this suggests that the post-Newtonian corrections might have an impact on the estimate of the mass density. 

Subsequently, we analysed the problem in the framework of the exact solutions of Einstein's equations. In particular, we considered  neutral, stationary and  axisymmetric rotating dust as model for a disc galaxy, and described the mathematical properties of the corresponding exact solution. Then, in order to give a physical insight into this solution, we studied it using the so-called zero angular momentum observers (ZAMO), which are suitable to analyse physical processes in presence of the symmetries considered. Using this formalism, we defined some useful observables, such as the dust velocity and the conserved quantities; in addition we  expressed the space-time metric exploiting the corresponding orthonormal tetrad. However, actual measurements on far away galaxies are not performed by the ZAMO but by asymptotic inertial observers who measure frequency shifts: to this end, we calculated the exact relations for the frequency shift of light, which can be used to explore the physical content of the exact solution.

We obtained further insight into the exact solution by considering some low energy limits, thanks to an expansion in (negative) powers of $c$. Besides the trivial special relativistic limit, we investigated the Newtonian limit and what we called the ``strong gravitomagnetic limit'' (SGM): notice that in this limit, which is naturally obtained from the exact solution, dragging effects are of the same order as Newtonian ones.
In particular, we showed that in the Newtonian limit a flat velocity profile can be achieved only on the basis of unphysical constraints on the mass distribution, whose density should vanish in the flat region. Things are quite different in the SGM limit, where dragging effects have an impact on the density profile required to match the flat velocity profile, in agreement with the analysis performed in the weak-field limit using the gravitomagnetic analogy.   In addition, using the frequency shift analysis, we showed that a flat velocity profile cannot be obtained in the Newtonian limit, while it naturally emerges when dragging effects are taken into account. As a particular case, we discussed the solution proposed by \citet{Balasin:2006cg}, which corresponds to a rigid rotation of the dust, and we pointed out some unphysical features: for instance, in this solution the redshift as seen from an inertial observer is always linearly increasing, while for an observer co-rotating with the galaxy the redshift is zero (see also \citet{Astesiano:2022gph}). It is worth remarking that even though the results we discussed refer to galaxies, the same system of equations can provide a good description of other self-gravitating system in the universe, such as cluster of galaxies.

{Our theoretical analysis, which encompasses both the weak-field limit of GR and exact solutions, shows that dragging effects may have a relevant impact in understanding galactic dynamics, due to the fact that they introduce an additional degree of freedom with respect to the  Newtonian case. In particular, our work provides a theoretical background to the recent publications by \citet{ludwig2021galactic,crosta2020testing} where models based on dragging effects were successfully used to fit data coming from galaxies rotation curves.}

Accordingly, we suggest that a better understanding of the mass content can be achieved using this approach, which might shed new light on the role of the dark matter, whose origin can partly be geometric.

\appendix

\newpage

\section{Exact solution  for rotating dust } \label{Comp}
Here, we make a comparison with the notation used  in the book by \citet{stephani_kramer_maccallum_hoenselaers_herlt_2003}  (see pages 330-333), where the general class of exact solutions is given in terms of different functions. We give the explicit map. Using their notations, after defining
\begin{align}
    \beta_{a} \equiv \frac{H_a}{H\eta},
\end{align}
they write down the following two last equations of motion
\begin{align}
    W^{-1}\left[(\beta W)_{,a}+(\frac{H}{\eta})(\frac{\eta^2}{H})_{,a} \right]= \epsilon_{ab} \gamma^{,b} \rightarrow \Delta \gamma&=0, \label{Gammas}\\
    D^{a}W_{,a}&=0,
\end{align}
where 
\begin{align}
    \Delta \gamma= \gamma_{rr}+ \frac{\gamma_r}{r}+ \gamma_{zz}.
\end{align}
For the second equation they state that it is always possible to choose $W=r$.\\
They claim that the full solution is given by choosing a function $\gamma$ and an axysimmetric solution of $\Delta \gamma=0$. Once $\eta(H)$ and $\gamma(r,z)$ are given, one obtain the function
\begin{align}
    2 \eta+ r^2 \beta- \int \frac{\eta}{H} dH, \label{GF}
\end{align}
from Eq. (\ref{Gammas}) and consequently $H(r,z)$ and finally 
\begin{align}
    H_{a}= 2 \eta \Omega_{,a},
\end{align}
which we wrote in Eq. (\ref{eta cond}). From our perspective the function $(\ref{GF})$ is exactly the function $\mathcal{F}$, which is related to $\gamma$ trough Eq. (\ref{Gammas}), or explicitly as
\begin{align}
   \frac{\mathcal{F}_{,r}}{r}= \gamma_{,z}, \quad \frac{\mathcal{F}_{,z}}{r}= -\gamma_{,r}.
\end{align}
Substituting these relations into $\Delta \gamma=0$ automatically solves the equations, but the closure of the form gives the consistency equation
\begin{align}
    \gamma_{,zr}-\gamma_{,rz}=0, \rightarrow \mathcal{F}_{,rr}- \frac{1}{r} \mathcal{F}_{,r}+ \mathcal{F}_{,zz}=0,
\end{align}
which is exactly the equation of motion (\ref{harm}).

\section{More details on the SGM limit} \label{APP}

The SGM limit is not equivalent to the standard gravitomagnetic approach discussed in Section \ref{Gravitomagn}, but it can be seen as a strong version of it. A simple inspection of the SGM metric 
\begin{align}
    ds^2= -c^2\left(1-\frac{2U}{c^2}-\frac{a^2}{c^2r^2}\right) dt^2 -2 a dt d\phi+ r^2\left(1+ \frac{2U}{c^2}\right) d\phi^2+e^{\Psi}\left(dr^2+dz^2\right),
\end{align}
and the gravitomagnetic one in the axisymmetric case
\begin{align}
\mathrm{d} s^2&= -c^2 \left(1-2\frac{\Phi}{c^2}\right) dt^2 -\frac4c A d\phi dt + \left(1+2\frac{\Phi}{c^2}\right)\delta_{ij}dx^i dx^j, \label{gravaxi1}
\end{align}
shows the substantial difference: the $g_{t\phi}$ term is of order $c^{0}$ in the former and $c^{-1}$ in the latter. In fact, assuming a form of $H$ as in the dragging limit, and setting $a \rightarrow 2  A/c$ in $\chi$, it is possible to obtain the standard gravitomagnetism at the leading order from the general system of equations. This is the reason why we introduced the term ``strong gravitomagnetism". In fact, let us check that the with the substitution $a \rightarrow 2 A/c$ the density is given by Eq. (\ref{deltadens}), and that  Eq. (\ref{eq:posissondelta}) is obtained at the leading order. Remembering that $b= r^{-2}a$ and after the substitution $a \rightarrow 2 A/c$,  Eq. (\ref{ddz0}) on the galactic plane ($z=0$) becomes
\begin{align}
  \frac{8\pi G}{c^2} \rho(z=0)=\frac{4v}{c^2} \frac{v_{,r}}{r} - \frac{4}{c^3} \left(2\frac{ A}{r^3}-\frac{ A_{,r}}{r^2}\right) \left(v-rv_{,r}\right)+O(c^{-4}).
\end{align}
The velocity $v$ used in the second part of the work is the velocity with respect to the ZAMO, while in the first part it is the coordinate velocity $v=r \Omega$. Therefore we must send $v \rightarrow v- \frac{2 A}{cr}$ to obtain
\begin{align}
    \frac{8\pi G}{c^2}\rho(z=0)=\frac{4v}{c^2} \frac{v_{,r}}{r}-\frac{4}{c^3}\frac{A_r}{r^2} \left(v+rv_{,r}\right)+O(c^{-4}),
\end{align}
which means
\begin{align}
    \frac{8\pi G}{c^2} \delta\rho=-\frac{4}{c^3}\frac{A_r}{r^2} \left(v+rv_{,r}\right)+O(c^{-4}). \label{dens1}
\end{align}
Let us check that this coincides with the result of the application of Eq. (\ref{eq:posissondelta}) in the axisymmetric case. Taking in account Eq. (\ref{gravaxi1}), we make use of the usual cylindrical vector basis $\mb u_{r}, \mb u_{\varphi}, \mb u_{z}$ to write
\begin{align}
 \mathbf{A}= \frac{A}{r}\mb u_{\varphi},\quad \mathbf{v}= v \mb u_{\varphi}.
\end{align}
These fields have the following rotors
\begin{align}
    \mathbf{B}=& \nabla \times \mathbf{A}=   \nabla (\frac{A}{r}) \times \mb u_{\varphi}+ \frac{A}{r} \nabla \times \mb u_{\varphi}=\frac{A_{,r}}{r} \mb u_{z}- \frac{A_{,z}}{r} \mb u_{r}, \\
    \bm \omega =& \nabla \times \mathbf{v}= \left(rv_{,r}+v\right) \frac{\mb u_{z}}{r}-v_{,z} \mb u_{r},
\end{align}
where we used the known fact $\nabla \times \mb u_{\varphi}= r^{-1} \mb u_{z}$. Eventually, we get
\begin{align}
    \delta \rho(z=0)=-\frac{1}{2\pi G c}\mb B \cdot \bm \omega= - \frac{1}{2\pi G c} \frac{A_r}{r^2} \left(v+r v_{,r}\right)+O(c^{-2}),
\end{align}
which coincides with Eq. (\ref{dens1}).

\begin{acknowledgments}
The authors thank Antonello Ortolan,  Clive C. Speake and Federico Re  for stimulating discussions on this topic.
\end{acknowledgments}

\bibliography{GEM_fluid}

\end{document}